\documentclass[aps,pra,twocolumn,groupedaddress]{revtex4}

\usepackage{graphicx}

\begin{document}

\title{Coherent response to optical excitation\\in a rare-earth ion doped crystal}
\author{J. Ruggiero, T. Chaneli\`ere, J.-L. Le Gou\"et}
\email[]{jean-louis.legouet@lac.u-psud.fr}
\affiliation{Laboratoire Aim\'e Cotton, b\^atiment 505, campus universitaire, 91405 Orsay cedex, France}

\date{\today}

\begin{abstract}
We investigate coherent propagation through a large optical density Tm$^{3+}$:YAG crystal. Using an ultra-stable laser, fiber filtering and site selection, we investigate the transmitted pulse temporal profile. The plane wave condition is satisfied by selection of the illuminated spot central area. We pay special attention to $\pi$-pulse transmission in the prospect of implementing optical quantum storage protocols.  
\end{abstract}


\maketitle

\section{Introduction}
Coherent response to optical excitation has been investigated for a long time. Actually, mere linear absorption of an optical beam through a material slab represents the most elementary manifestation of this feature. Indeed the transmitted beam results from the coherent combination of the incident beam and the instantaneous material response. It has also been realized early that coherent optical response is related to the collective quantum excitation of an atomic ensemble. As pointed out by R. Dicke~\cite{dicke1954}, {\it all the molecules are interacting with a common radiation field and hence cannot be treated as independent}. However, a semiclassical treatment is generally adequate to describe the coherent response ~\cite{dicke1954}, especially if the number of atoms per $\lambda$-cube is small enough so that superfluorescence cannot start up on the experiment timescale ~\cite{rehler1971}. In the framework of the semiclassical description, McCall and Hahn were able to derive an area theorem that accounts for the propagation of an arbitrary intensity pulse through an arbitrary optical density slab~\cite{mccall1967}, giving rise to the notion of Self-Induced Transparency (SIT). To be coherent, the process must take place on a time scale much shorter than the atomic superposition state lifetime. Experimental demonstration in the nanosecond range was performed at once in both solids and atomic vapors at room temperature ~\cite{mccall1967,slusher1972}. The McCall and Hahn theorem only predicts the pulse area behavior. For a more detailed description, especially to investigate the pulse deformation, one has to solve the Maxwell-Bloch (MB) equation numerically. 

The connection with solitons have stimulated the interest in SIT. It was shown~\cite{mccall1969} that a $2\pi$-area pulse with hyperbolic secant temporal profile can propagate without alteration through an absorbing medium. The breakdown of large area pulses into $2\pi$-area components was also predicted~\cite{mccall1969} and observed~\cite{slusher1972}. A couple of decades later, the discovery of Electromagnetically Induced Transparency (EIT) in ensembles of three-level atoms~\cite{boller1991} renewed the interest in the coherent response of optically thick media. Actually a variety of induced transparency processes have been investigated in three-level systems~\cite{konopnicki1981,harris1993}. The interplay of EIT with SIT was also considered~\cite{kozlov2000}. In recent years, effort has been directed to the ultra-short pulse regime, where the usual assumptions, such as the rotating wave and slowly varying envelope approximations may not be valid ~\cite{schlottau2005}.   
  
Rare-earth ion doped crystals (REIC) represent an excellent testbed for coherent optical response investigation. Most of the limitations pointed out by Slusher in his review paper on self-induced transparency ~\cite{slusher1974} are relaxed in these materials. The superposition state lifetime may reach hundred of microseconds and is not limited by a transit time across the light beams. No diffusive motion can affect the spatial phase of the atomic states. The inhomogeneous broadening can be considered as infinite. Generally, the level degeneracy is completely lifted by the crystal field, so that the ions behave as true two-level atoms on the propagation timescale. Other features depend on the specific host matrix. In Yttrium Aluminium Garnet (YAG, Y$_3$Al$_5$O$_{12}$ ) for instance, the $D_2$ symmetry of substitution sites forbids the existence of permanent dipole moments. As a consequence, relaxation processes such as optical excitation induced instantaneous spectral diffusion (IST) are strongly reduced. In this matrix, appropriate crystal cutting and adequate polarization orientation make the ions in different sites interact with the same dipole moment projection on the incoming field~\cite{sun2000}. 

Despite of such attractive features, very few SIT coherent propagation data have been collected in REIC \cite{zumofen1999,greiner2000}. Ref~\cite{greiner2000} reports on experimental work in Tm$^{3+}$:YAG. However coherent excitation conditions are not clearly satisfied since the laser is not stabilized. In addition inserting the sample in a cavity makes the quantitative analysis more difficult. In Ref~\cite{zumofen1999}, propagation is investigated in Pr$^{3+}$:YSO. The laser is properly stabilized but an unexpected low transmission factor is observed.  

In the present paper our aim is twofold. First we propose an original solution of the MB equation that analytically accounts for the instantaneous part of the atomic response. This fast component spreads over a broad spectral interval and complicates the numerical solution when the absorption line is strongly inhomogeneously broadened. Second, we present a detailed experimental investigation of coherent pulse propagation in a large optical density Tm$^{3+}$:YAG crystal at liquid helium temperature. Our work is related to topical research on quantum storage in REIC. The storage protocols generally involve a preparation step that can require sophisticated excitation procedures~\cite{pryde2000,deseze2003,deseze2005}. We presently concentrate on single rectangular pulse propagation, as a preliminary step to the investigation of more complex pulse sequences~\cite{pryde2000}. In the context of quantum storage, we are more interested in the spectral and spatial distribution of atomic excitation than in SIT by itself. The $\pi$-pulse case deserves special attention since all the resonantly excited ions should undergo $\pi$-pulse excitation, wherever they sit in the sample depth. 

The paper is organized as follows. In Section 2 we detail the solution of the MB equation. In section 3 we discuss the features of rectangular pulse simulated propagation. The setup is described in Section 4. We present and discuss the experimental data in section 5. 

\section{the instantaneous response}
In this section we rely on physical arguments in order to simplify the computation procedure. Specifically we show that one can simply account for the contribution from atoms far from resonance. This way the sum over the inhomogeneous width can be limited to a narrower spectral interval.

Let us consider a spatially and spectrally uniform distribution of two-level atoms. Spectral uniformity means that the inhomogeneous width of the optical transition is infinite. Let a monochromatic plane wave be directed to the sample. At the input side $z=0$ the electric field E(z,t) reads as:
\begin{equation}
\mathrm{E}(0,t)=\mathrm{A}(0,t)\cos (\omega_Lt+\phi)
\end{equation}
where $\phi$ is time-independent. Within the frame of the slowly varying amplitude (SVA) and rotating wave (RWA) approximations, the Maxwell Bloch equations reads as:
\begin{equation}\label{MB}
\begin{array}{ll}
\partial_z\Omega(z,t)&=-\displaystyle\frac{\alpha}{2\pi}\int d\omega_{ab}\mathrm{v}(\omega_{ab};z,t)
\\[0.3cm]
\partial_t\mathrm{u}(\omega_{ab};z,t)&=-\Delta \mathrm{v}(\omega_{ab};z,t)\\[0.3cm]
\partial_t\mathrm{v}(\omega_{ab};z,t)&=-\Omega(z,t)\mathrm{w}(\omega_{ab};z,t)+\Delta \mathrm{u}(\omega_{ab};z,t)\\[0.3cm]
\partial_t\mathrm{w}(\omega_{ab};z,t)&=\Omega(z,t)\mathrm{v}(\omega_{ab};z,t)
\end{array}
\end{equation}
where $\Delta=\omega_{ab}-\omega_L$ and where $\Omega(z,t)$, u($\omega_{ab};z,t$), v($\omega_{ab};z,t$) and w($\omega_{ab};z,t$) respectively represent the Rabi frequency and the components of the Bloch vector B($\omega_{ab};z,t$). Let $L$ represent the sample thickness. To further simplify the problem, we assume that the incoming pulse evolution characteristic time, i.e. the inverse spectral width $1/\Delta_p$, is much longer than $L$/c. Therefore we can neglect the time derivative $\partial_t\Omega(z,t)/\mathrm{c}$ in the wave equation.  

To solve the equations numerically, one would like to get the spatial distribution $\Omega(z,t)$ at time $t$ in terms of $\Omega(0,t)$, the boundary value at the sample input, and of B($\omega_{ab};z,t-\tau$), where $\tau \Delta_p<<1$. Then one would solve the Bloch equation at time $t$, starting with B($\omega_{ab};z,t-\tau$) and $\Omega(z,t-\tau)$. It should be stressed that the material response at time $t$ cannot be simply expressed in terms of the Bloch vector at $t-\tau$. Indeed, due to the infinite inhomogeneous bandwidth, the material gives rise to an instantaneous response. Let us formally solve the Bloch equation in the following way:
\begin{equation}\label{1} 
\begin{array}{ll}
\mathrm{v}(\omega_{ab};z,t)=\mathrm{v}(\omega_{ab};z,t-\tau)\cos(\Delta\tau)\\[0.3cm]
+\mathrm{u}(\omega_{ab};z,t-\tau)\sin(\Delta\tau)\\[0.3cm]
-\displaystyle\int_0^{\tau}\Omega(z,t-\tau')\mathrm{w}(\omega_{ab};z,t-\tau')\cos(\Delta\tau')d\tau'
\end{array}
\end{equation}
The first two terms on the right-hand side corresponds to the free precession of the Bloch vector from $t-\tau$ to $t$. The last term reflects the coupling to the field during this time interval. Initially $\mathrm{w}(\omega_{ab};z,t_0)=-1$, since all the atoms sit in the ground state. As the pulse propagates through the sample, $\mathrm{w}(\omega_{ab};z,t)+1$ departs from $0$ over a spectral interval given by the pulse width $\Delta_p$. However, according to Eq. \ref{1}, the driving field uniformly excites the atoms all over the inhomogeneous width since the weight factor $\mathrm{w}(\omega_{ab};z,t_0)$ is initially close to $-1$ everywhere. Summing over an infinite inhomogeneous width clearly raises a numerical computation issue. An infinite width contribution also reflects an instantaneous response feature that we intend to express analytically, thus relaxing the computation problem. Substituting $\mathrm{v}(\omega_{ab};z,t)$ into the wave equation one obtains:
\begin{equation}
\begin{array}{ll}
\partial_z\Omega(z,t)=-\displaystyle\frac{\alpha}{2\pi}\int d\omega_{ab}\int_0^{\tau}\Omega(z,t-\tau')\cos(\Delta\tau')d\tau'\\[0.5cm]
+\displaystyle\frac{\alpha}{2\pi}\int d\omega_{ab}\int_0^{\tau}\Omega(z,t-\tau')[\mathrm{w}(\omega_{ab};z,t)+1]\cos(\Delta\tau')d\tau'\\[0.5cm]
-\displaystyle\frac{\alpha}{2\pi}\int d\omega_{ab}\mathrm{v}_{free}(\omega_{ab};z,t-\tau,t)
\end{array}
\end{equation}   
where:
\begin{equation}
\begin{array}{ll}
\mathrm{v}_{free}(\omega_{ab};z,t-\tau,t)=\\[0.3cm]
\mathrm{v}(\omega_{ab};z,t-\tau)\cos(\Delta\tau)+\mathrm{u}(\omega_{ab};z,t-\tau)\sin(\Delta\tau)
\end{array}
\end{equation} 
The instantaneous response contribution has been isolated in the first term on the right-hand side. This easily reduces to $-\frac{\alpha}{2\pi}\Omega(z,t)$. The second term, of order $-\frac{\alpha}{2}\Delta_p\tau\Omega(z,t)$, can be neglected since $\Delta_p\tau<<1$. Finally the wave equation solution reads as:
\begin{equation}\label{2}
\begin{array}{ll}
\Omega(z,t)=\Omega(0,t)\mathrm{e}^{-\frac{\alpha}{2}z}\\[0.3cm]
-\displaystyle\frac{\alpha}{2\pi}\int_0^z dz'\mathrm{e}^{-\frac{\alpha}{2}(z-z')} \int d\omega_{ab}\mathrm{v}_{free}(\omega_{ab};z',t-\tau,t)
\end{array}
\end{equation}
In the small area limit the second term on the right-hand side can be neglected. The equation then reduces to the Bouguer law of absorption ~\cite{bouguer1760}. As a starting point for numerical computation, the equation conveniently expresses $\Omega(z,t)$ in terms of $\Omega(0,t)$ and of B($\omega_{ab};z,t-\tau$). In order to get B($\omega_{ab};z,t$) in terms of B($\omega_{ab};z,t-\tau$) and $\Omega(z,t-\tau)$ one complements Eq. \ref{1} with the integral solutions for u($\omega_{ab};z,t$) and w($\omega_{ab};z,t$): 
\begin{equation}\label{3} 
\begin{array}{ll}
\mathrm{u}(\omega_{ab};z,t)&=\mathrm{u}_{free}(\omega_{ab};z,t-\tau,t)\\
&+\displaystyle\int_0^{\tau}\Omega(z,t-\tau')\mathrm{w}(\omega_{ab};z,t-\tau')\sin(\Delta\tau')d\tau'\\
\mathrm{w}(\omega_{ab};z,t)&=\mathrm{w}(\omega_{ab};z,t-\tau)\\
&+\displaystyle\int_0^{\tau}\Omega(z,t-\tau')\mathrm{v}(\omega_{ab};z,t-\tau')d\tau'\
\end{array}
\end{equation}
where:
\begin{equation}
\begin{array}{ll}
\mathrm{u}_{free}(\omega_{ab};z,t-\tau,t)=\\[0.3cm]
\mathrm{u}(\omega_{ab};z,t-\tau)\cos(\Delta\tau)-\mathrm{v}(\omega_{ab};z,t-\tau)\sin(\Delta\tau)
\end{array}
\end{equation} 

According to Eq. \ref{3}, the second term on the right-hand side of Eq. \ref{2} still apparently contains far-from-resonance contributions. Indeed u($\omega_{ab};z,t$) and v($\omega_{ab};z,t$) are built from w($\omega_{ab};z,t$) all over the infinite inhomogeneous width. To show the instantaneous response is actually contained in the first term, let us define the new variables: 
\begin{equation} 
\begin{array}{ll}
\mathrm{U}(\omega_{ab};z,t)=\\
\mathrm{u}(\omega_{ab};z,t)+\displaystyle\int_0^\infty\Omega(z,t-\tau')\sin(\Delta\tau')d\tau'\\
\mathrm{V}(\omega_{ab};z,t)=\\
\mathrm{v}(\omega_{ab};z,t)-\displaystyle\int_0^\infty\Omega(z,t-\tau')\cos(\Delta\tau')d\tau'
\end{array}
\end{equation}
Those new variables vanish far from resonance with the driving field. Then one easily verifies that Eq. \ref{2} is left unchanged if u($\omega_{ab};z,t-\tau$) and v($\omega_{ab};z,t-\tau$) are respectively replaced by U($\omega_{ab};z,t-\tau$) and V($\omega_{ab};z,t-\tau$). 

\section{pulse distorsion and area theorem}
\begin{figure}
\includegraphics[width=8cm]{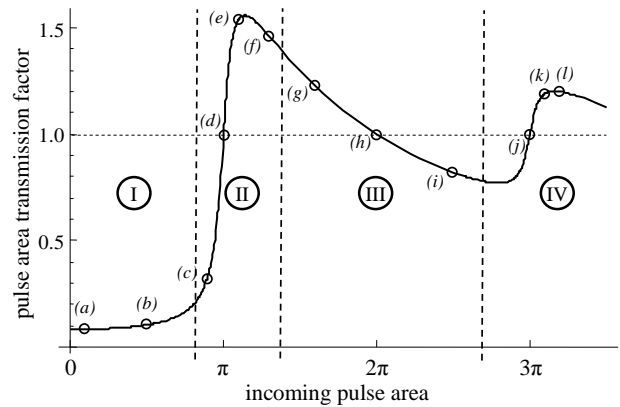}
\caption{area theorem: pulse area transmission factor $A_\mathrm{out}/A_\mathrm{in}$ as a function of the incoming pulse area $A_\mathrm{in}$. Labels ({\it a}) to ({\it l}) refer to the different input-area conditions considered in Fig.~\ref{fig_simul}.}   
\label{fig_area}
\end{figure}

\begin{figure}
\includegraphics[width=8cm]{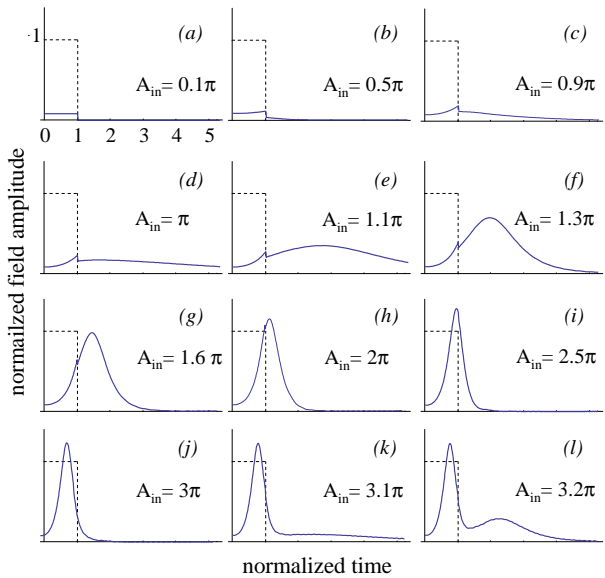}
\caption{temporal distorsion of an incoming rectangular pulse after propagation through a $\alpha L=5$ sample, for different initial area values. Time and outgoing pulse amplitude are respectively normalized to the duration and the amplitude of the incoming pulse.}
\label{fig_simul}
\end{figure}

According to the McCall and Hahn theorem~\cite{mccall1967}, the transmitted pulse area, $A_\mathrm{out}$, can be expressed in terms of the incoming pulse, $A_\mathrm{in}$, as:
\begin{equation} 
\tan(A_\mathrm{out}/2)=\mathrm{e}^{-\alpha L/2}\tan(A_\mathrm{in}/2)
\end{equation}
where the pulse area is defined as $A=\int\Omega(t)\mathrm{d}t$. The corresponding variation of the transmission factor $A_\mathrm{out}/A_\mathrm{in}$ as a function of $A_\mathrm{in}$ is displayed in Fig. \ref{fig_area}. The shape distorsion of an incoming rectangular pulse is presented in Fig. \ref{fig_simul} for different pulse area values. The box labels help to locate the propagation regime in Fig.\ref{fig_area}. The distorsion can be understood in the light of the area theorem and of energy conservation. Four different regions can be identified in Fig.\ref{fig_area}. In region (I) nearly no light is emitted after the incoming pulse extinction. The pulse simply obeys the Bouguer law and is hardly distorted. In region (II), located around $\pi$ area, the pulse stretches in order to comply with two contradictory prescriptions. On the one hand the energy absorption increases, since all the resonantly excited atoms are promoted to the upper level, at any depth within the sample. On the other hand the area transmission factor grows larger that unity. The region (III) is centered on $2\pi$ area. The transmission factor is still close to unity but less energy is absorbed since the resonantly excited atoms are returned to their ground state. In this region the outgoing pulse duration reduces, getting closer to that of the incoming pulse. In this region the pulse tends to the expected soliton hyperbolic secant shape. Finally the pulse area undergoes another increase in region (IV), around $3\pi$ area. As in region (II), this area increase conflicts with the energy depletion by the resonantly excited atoms that are left in the upper level. This apparent contradiction is solved by the emergence of a stretched secondary component in the tail of the main pulse. However, the $3\pi$ pulse is less distorted than the $\pi$ pulse by the increase of energy transfer to the atoms. Indeed, to excite the same number of atoms, a $3\pi$ pulse consumes a $9$ times smaller fraction of the available energy than a $\pi$ pulse.

The rapid growing of the long tails, in both the $\pi$ and $3\pi$ regions, provides us with key features for experimental data analysis.   

\section{Experimental setup}
The $0.5\%$ at. Tm$^{3+}$:YAG crystal is cooled down to $\cong2.2$K in a helium bath cryostat. The experiment is performed on the $^3$H$_4 - ^3$H$_6$ transition at 793nm. The sample length is $L=5$mm. At the operating temperature the opacity is measured to be $\alpha L\cong 5$. The crystal sides are cut perpendicular to direction [1, -1, 0], and the light beam is polarized along direction [1, 1, 1]. This way, three sites out of six interact with the driving field, and they do so with the same Rabi frequency.
 
\begin{figure}
\includegraphics[width=7cm]{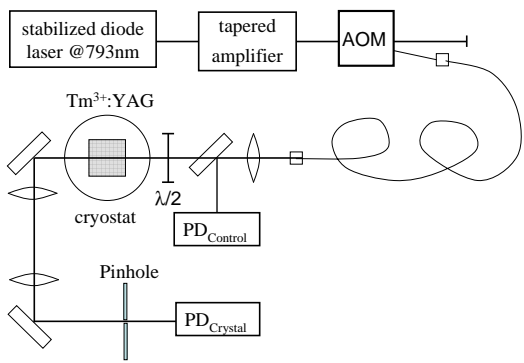}
\caption{experimental setup.}
\label{fig_setup}
\end{figure}

The sample is excited by a monochromatic semiconductor laser (Fig.\ref{fig_setup}). We carefully control the spectro-temporal and spatial properties of the source. The laser linewidth is first reduced to less than 1kHz by Pound-Drever-Hall locking to a high-finesse Fabry-Perot cavity. After boost up through a tapered semiconductor amplifier, the laser beam is precisely temporally shaped by an acousto-optic modulator (AOM). The AOM is directly driven by a high frequency arbitrary wave form generator (Tektronix AWG520), operating at 1Gs/s, which guarantees precise amplitude and phase control. The beam is then spatially filtered by a 2m-long single mode optical fiber. 

The experimental conditions must be consistent with the plane wave assumption of the theory. First we make the laser beam depth of field (DF) $>>L$. The DF, defined as twice the Rayleigh range, is given by $2\pi nw_0^2/\lambda$, where $n=1.82$ stands for the YAG index of refraction and where the waist has been adjusted to $w_0\cong25\mu$m. Therefore DF$\cong9$mm, which is significantly larger than $L$. In order to put the beam waist at the crystal center, we position the focussing lens at maximum Rabi frequency, as probed by an optical nutation signal. Then we use a pin-hole to select the signal emerging from the center of the illuminated spot. The sample is imaged on the $50\mu$m-pinhole through a telescope with a $\times4$ magnification factor. This corresponds to $8\%$ peak-to-peak excitation intensity variation over the detected area. The signal is detected on an avalanche photodiode.

The input intensity is monitored by photodetector PD$_\mathrm{control}$(Fig.\ref{fig_setup}). We again resort to optical nutation measurements for calibrating the detector in Rabi frequency units. Rabi frequency can be precisely deduced from the optical nutation signal in the low absorption limit only. Because of the large sample opacity at resonance, we have to detune the laser about $0.1$nm from resonance to perform the calibration. The transmitted intensity is detected on PD$_\mathrm{crystal}$. We calibrate this detector with the raising edge of the rectangular pulse. Indeed, immediately after switch-on, the transmitted intensity simply reads as $I_0\mathrm{exp}(-\alpha L)$, where $I_0$ represents the input pulse intensity that is measured on PD$_\mathrm{control}$.

\section{results and discussion}
We study the propagation of rectangular-shaped pulses. The pulse duration is adjusted to $7\mu$s, which is significantly smaller than the atomic coherence lifetime of $\cong50\mu$s.

\begin{figure}
\includegraphics[width=8cm]{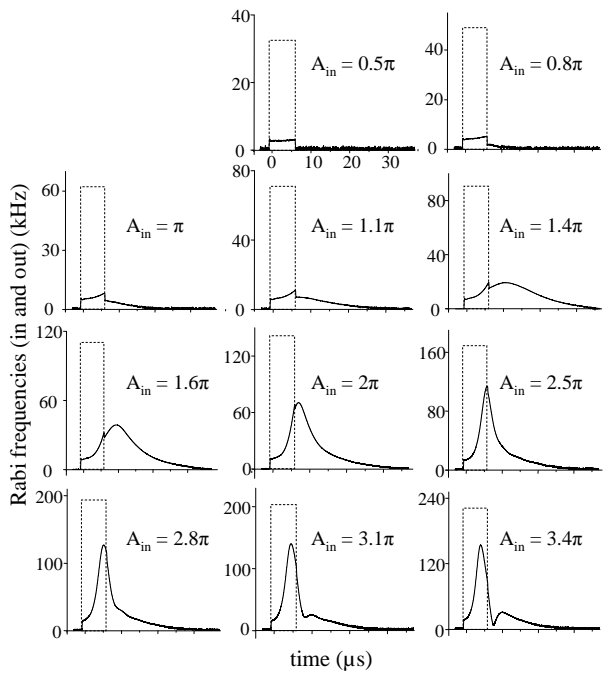}
\caption{Experimental profile of the transmitted pulse amplitude (solid line). Input area (A$_{in}$) ranges from $0.5\pi$ to $3.4\pi$. Input pulses (dotted line) are rectangular with 7$\mu$s-duration.}
\label{fig_profils_exp}
\end{figure}

The input area ranges from $0.5\pi$ to $3.4\pi$. The observed temporal profiles are displayed in Fig. \ref{fig_profils_exp}. As expected, the pulse is strongly stretched around input $\pi$ area, since the area has to be conserved despite of large resonant absorption. Then, as the input area is increased, the pulse shrinks back, which is consistent with reduced absorption. Finally, around $3\pi$, the pulse spreads again as resonant absorption increases, growing a side lobe to simultaneously satisfy area conservation and energy drop. The expected output area evolution is also observed. However, agreement with theory is only qualitative. 

\begin{figure}
\includegraphics[width=8cm]{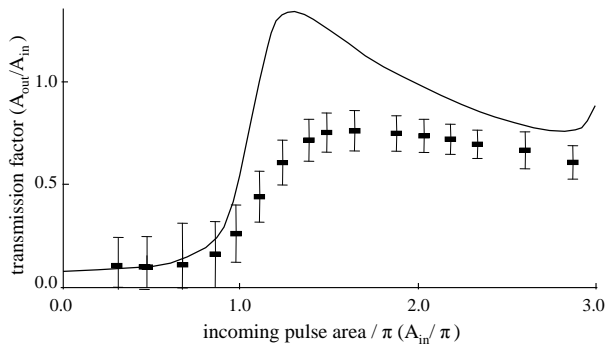}
\caption{experimental and theoretical pulse area transmission factor $A_\mathrm{out}/A_\mathrm{in}$ as a function of the incoming pulse area $A_\mathrm{in}$. Input pulses are rectangular with 7$\mu$s-duration. Theory accounts for a finite coherence lifetime of 50$\mu$s}
\label{fig_area_exp}
\end{figure}

As seen in Fig. \ref{fig_area_exp} the ratio of the output area to the input area lags behind the predicted value, especially in the region of $\pi$ and $3\pi$ incoming pulses. The theoretical profile in Fig. \ref{fig_area_exp} accounts for the finite dipole lifetime that has been set to $50\mu s$. As compared with Fig. \ref{fig_area}, the finite coherence lifetime slightly reduces the computed transmission factor, especially in the $\pi$ and $3\pi$ regions where the propagating pulse is strongly stretched. However this reduction is not sufficient to fit the experiment. Finally the excess of absorption reminds us of previously reported data ~\cite{zumofen1999}, although agreement with theory has been improved in our experiment. A possible fault of the setup is the absence of antireflection coating on the crystal. As a consequence a reflected field counterpropagates through the excited medium, with an amplitude reflection coefficient as large as $30\%$, given the high index of refraction of YAG. One may also fear that the plane wave assumption breaks down in the $\pi$-pulse region. Indeed, the portions of the Gaussian beam profile corresponding to areas smaller than $\pi$ are absorbed more efficiently than the central part of the beam~\cite{slusher1974}. Such a stripping of the outer portions of the beam tends to reduce the waist and to move it to the output of the sample. The effective depth of field gets rapidly smaller than the sample length, since it varies as the square of the waist. Only a 3D Maxwell equation would correctly account for the resulting focussing and diffraction effects.      

\section{conclusion}
The propagation of rectangular pulses has been investigated in monochromatic plane wave conditions. In order to consistently account for the very broad inhomogeneous bandwidth of the absorption line, the standard MB theory has been adjusted so that the instantaneous radiative response has been described analytically. Experimental data qualitatively agree with theory. However a significant quantitative discrepancy subsists, especially in the most interesting $\pi$-pulse region, where the resonantly excited Bloch vector should undergo a $\pi$ rotation at any depth inside the sample ~\cite{ruggiero2008}. In the context of quantum optical storage, the need for efficient preparation of large optical density sample has been stressed recently ~\cite{afzelius2008} and should stimulate further coherent propagation investigation. Extension to three-level $\Lambda$-systems might be appropriate.

\end{document}